\documentclass[prl,letterpaper,aps,10pt,superscriptaddress,twocolumn,floatfix,showpacs]{revtex4-1}
\usepackage{graphicx}
\usepackage{amsmath}
\usepackage{amsfonts}
\usepackage{amssymb}
\usepackage{epsfig}
\usepackage[pdftex]{color}
\usepackage{amsmath,graphicx,amssymb,xcolor,subfigure,upgreek}
\usepackage[colorlinks, linkcolor=blue, citecolor=blue, urlcolor=blue, breaklinks=true]{hyperref}
\usepackage{bbm}
\usepackage{color}

\def\g2R{g^{(2)}_R}

\def\w0{w_0}
\def\avg#1{\mathinner{\langle{#1}\rangle}}

\def\rb{{\bf r}}

\def\hge{\hat{\sigma}^{ge}}  
\def\heg{\hat{\sigma}^{eg}} 
\def\hee{\hat{\sigma}^{ee}}  
\def\het{\hat{\sigma}^{et}}
\def\hte{\hat{\sigma}^{te}} 
\def\hee{\hat{\sigma}^{ee}}

\def\ket#1{\mathinner{|{#1}\rangle}}

\newcommand{\Braket}[2]{\langle{#1}|{#2}\rangle}

\def\Heff{\hat{H}_{\rm eff}}
\def\sabs{\sigma_{\rm abs}}

\newcommand{\fref}[1]{Fig.~\ref{#1}}

\def\kb{\bf k}
\def\hR{\hat{R}}
\def\sabsinc{\sigma_{\rm abs, inc}^{\rm single}}
\def\sabsc{\sigma_{\rm abs, c}^{\rm single}}

\newcommand\varpm{\mathbin{\vcenter{\hbox{%
  \oalign{\hfil$\scriptstyle+$\hfil\cr
          \noalign{\kern-.3ex}
          $\scriptscriptstyle({-})$\cr}%
}}}}
\newcommand\varmp{\mathbin{\vcenter{\hbox{%
  \oalign{$\scriptstyle({+})$\cr
          \noalign{\kern-.3ex}
          \hfil$\scriptscriptstyle-$\hfil\cr}%
}}}}

\begin{document}
\title{Efficient frequency selective single photon antennas based on a bio-inspired nano-scale atomic ring design with 9-fold symmetry}
\author{Maria Moreno-Cardoner}
\affiliation{Institut f\"ur Theoretische Physik, Universit\"at Innsbruck, Technikerstr. 21a, A-6020 Innsbruck, Austria}
\author{Raphael Holzinger}
\affiliation{Institut f\"ur Theoretische Physik, Universit\"at Innsbruck, Technikerstr. 21a, A-6020 Innsbruck, Austria}
\author{Helmut Ritsch}
\affiliation{Institut f\"ur Theoretische Physik, Universit\"at Innsbruck, Technikerstr. 21a, A-6020 Innsbruck, Austria}
\date{\today}

\date{\today}
\begin{abstract}
Quantum emitters in confined arrays exhibit geometry dependent collective dynamics. In particular, nanoscopic regular polygon-shaped arrays can possess sub-radiant states with an exciton lifetime growing exponentially with emitter number. We show that by placing an extra resonant absorptive dipole at the ring center, such a structure becomes a highly efficient single-photon absorber with tailorable frequency. Interestingly, for exactly nine emitters in a nonagon, as it appears in a common biological light-harvesting complex (LHC2), we find a distinct minimum for its most dark state decay rate and a maximum of the effective absorption cross-section, surpassing that for a single absorptive emitter. The origin of this optimum for nine emitters can be geometrically traced to the fact that the sum of coupling strengths of a single ring emitter to all others including the center ring closely matches the coupling of the center to all ring emitters. The emerging dark collective eigenstate has dominant center occupation facilitating efficient energy absorption and fast transport. The resonance frequency can be tuned via ring size and dipole polarization. In analogy to parabolic antennas, the ring concentrates the incoming radiation at the center without being significantly excited, which minimizes transport loss and time.
\end{abstract}
\pacs{42.50.Ct, 42.50.Nn}
\maketitle
\section{Introduction}
Collective radiation effects such as sub-radiance and super-radiance \cite{dicke1954coherence,haroche1982superradiance,guerin2016subradiance,Solano2017superradiance} arising in tight ordered structures of dipole-dipole coupled quantum emitters are nowadays attracting renewed widespread interest \cite{temnov2005superradiance,zoubi2008bright,Porras2008collective,scully2009collective,Jenkins2012Controlled,Jenkins2013metamaterial,scully2015single,plankensteiner2015selective,bettles2015cooperative,Tudela2015Subwave,sutherland2016collective,bettles2016cooperative,bettles2016enhanced,shahmoon2016cooperativity,asenjo2016,asenjo2017exponential,ruostekoski2017arrays,hebenstreit2017subradiance,Chang2018colloquium,cottier2018role,guimond2019subradiant,PineiroOrioli2019dark,zhang2019theory, kornovan2019extremely,Zhang2020subradiant,Zhang2020universal,PineiroOrioli2020subradiance}. This is partially triggered by recent advances in implementing and controlling precise arrays of single quantum emitters at close distance, as e.g. in uniformly filled optical lattices \cite{rui2020subradiant,bakr2009quantumgas,sherson2010single,weitenberg2011single}, or structures of definable geometry using individually trapped atoms with optical tweezers \cite{barredo2016Atom,Endres2016,Barredo2018,Norcia2018,Schlosser2019Large}, and microwave superconducting q-bits \cite{wang2020controllable,mirhosseini2019Cavity,vanLoo2013Photon}. Such ordered dipole arrays create a novel quantum optics platform that enables an enhanced atom-light coupling which can outperform current bounds in quantum protocols including single photon storage \cite{jenkins2016many,asenjo2017exponential,manzoni2018optimization}, spectroscopy \cite{ostermann2014protected,henriet2019critical}, or optomechanics \cite{shahmoon2020cavity,shahmoon2020quantum}. Moreover they represent a genuine test bed to explore fundamental physics of quantum many body states of light and matter \cite{henriet2019critical,zhang2019theory,mahmoodian2019dynamics,perczel2020topological}. 

Among the different geometries, arrays forming a regular polygon with sub-wavelength inter-particle distance (from now on referred to as nanorings) show very intriguing radiative properties \cite{asenjo2017exponential,jen2018cooperative,olmos2019subradiance, moreno2019subradiance,holzinger2020nanoscale,cremer2020polarization}. On the one hand nanorings support extremely subradiant guided modes with a loss exponentially decreasing with the atom number \cite{asenjo2017exponential}, and can be used for an efficient excitation transport within a single \cite{olmos2019subradiance} or between two of these coupled rings \cite{moreno2019subradiance,cremer2020polarization}. On the other hand they possess collective eigenmodes whose field is confined in a subwavelength region at their geometrical center. For a proper choice of system parameters such nano-rings behave as optical resonators that can operate like a coherent laser source when pumping an additional emitter at the center \cite{holzinger2020nanoscale}. 

Here we show that in the weak driving regime such a ring structure can also behave like a parabolic mirror \cite{alber2013qed,Leuchs} concentrating the light at its focus and thus strongly enhancing the coupling of an incoming plane wave photon to the additional central emitter, where it can be absorbed, creating an effective absorption cross section way beyond that for a single free space atom. Surprisingly we find that a nano-ring of $N=9$ dipoles, as it appears in many common biological light harvesting complexes, called LHC2 \cite{bourne2019structure,cogdell2006architecture,mirkovic2017light,montemayor2018computational,caycedo2017quantum}, exhibits superior performance compared to other antenna atom numbers.  

To analyze this astonishing superiority of a nine-atom ring we first analyze its eigenmodes and find that there exists an extremely subradiant mode, whose central emitter excited state population is very large. Note that similar dark states can also appear for other number $N$ of ring emitters, but only when we precisely tune the dipole moment and transition frequency of the central emitter for each $N$. Despite being strongly subradiant, these states still couple to an incident plane wave coherent field. Tuned to the proper frequency, the high-Q resonance enhancement of the dark mode then creates a large steady state excitation of the central emitter strongly surpassing its free space excitation without the ring. Interestingly, throughout the process the ring atoms are only weakly excited in this process. It is important to note that the collection efficiency enhancement is even much larger if the incoming field does not directly illuminate the center absorber, but only couples to the antenna ring. 

Any additional irreversible decay occurring from the excited state of the central emitter will then allow to slowly but efficiently extract the collected energy. Hence the system can be regarded as a minimalist simulator of natural light-harvesting complexes present e.g. in purple bacteria that share a similar structure and optical properties  \cite{saer2017light,caycedo2017quantum,bourne2019structure}.

In general we will quantitatively show that the absorption spectrum of the central emitter is strongly modified by the presence of the outer ring and generally leads to a strongly enhanced maximal steady state absorption cross-section. The spectral properties can be tailored via the dipole moment and resonance energy of the central absorbing dipole. In essence, the outer ring, acts like a parabolic mirror or antenna, which redirects the incoming radiation towards the central emitter with very little intrinsic excitation. Hence radiative and other losses in the ring can be expected to be minimized \cite{caycedo2017quantum}.

\section{Model}
\begin{figure}[t]
\centering
\includegraphics[width=0.35\textwidth]{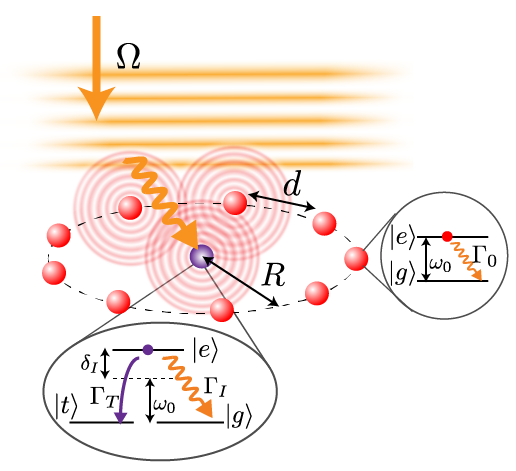}
\caption{Schematics of the system: a regular polygon of radius $R$ and distance $d$ of $N$ two-level emitters (antenna) is coupled via dipole-dipole interactions to a central emitter (impurity). The spontaneous emission rate of each ring emitter is given by $\Gamma_0$. The impurity can decay from the excited state (e) to the ground state (g) by spontaneous emission of a photon at rate $\Gamma_I$, or to an auxiliary state (t) via an additional irreversible channel at rate $\Gamma_T$. The impurity (g)-(e) transition frequency is detuned by $\delta_I$ from the antenna atoms. The whole system is weakly driven by a spatially coherent (temporally coherent or incoherent) external field.}
\label{Fig0}
\end{figure}

As a simple model of our antenna we use a regular polygon of $N$ equal point-like two-level emitters with dipole moment $\wp_i = \wp$ at fixed position, and an extra dipole at its center (referred to as ``impurity''). The distance between two neighbours in the ring and the radius are given by $d$ and $R$, respectively. The impurity possesses a transition with associated polarization $\wp_I = \sqrt{\Gamma_I} \wp$  detuned from other atoms transition frequency by $\delta_I$, and an additional decay channel into an auxiliary state (t) with associated rate $\Gamma_T$, to extract energy out of the system (see \fref{Fig0}). All the emitters (including the impurity) are interacting via vacuum mediated dipole-dipole interactions, which can be described (in the rotating frame with $\omega_0$) in the Born-Markov approximation by the master equation \cite{lehmberg1970radiation}  $\dot{\rho} = -i [ \hat{H} ,\rho ] + \mathcal{L}[\rho]$ ($\hbar = 1$). Here $\hat{H} = \sum_{i\neq j} J_{ij} \heg_i \hge_j$, and $\mathcal{L}[\rho] = \frac{1}{2} \sum_{i,j} \Gamma_{ij} \left( 2\hge_j \rho \heg_i - \heg_i \hge_j \rho - \rho \heg_i \hge_j \right)$, where $\hge_j$ ($\heg_j$) is the lowering (raising) operator between excited and ground state of emitter $j$. The dispersive and dissipative couplings are given respectively by $J_{ij} = \textrm{Re} \mathcal{G}_{ij}$ and $\Gamma_{ij}=-2\textrm{Im} \mathcal{G}_{ij}$, with $\mathcal{G}_{ij}$ proportional to the free space Green's tensor,
\begin{align}
\mathcal{G}_{ij} &= \frac{3\Gamma_0}{4 k_0^3 r^3 \wp^2} e^{i k_0 r} \boldsymbol{\wp}^{\alpha,*}_i\boldsymbol{\wp}^\beta_j \quad\times\\
&\left[\left(1 -i k_0 r - k_0^2 r^2\right) \delta_{\alpha \beta} 
 +\left(-3+3ik_0 r +k_0^2 r^2\right)\frac{\rb_\alpha \rb_\beta}{r^2} \right] \notag
\end{align}
(summation convention is used). Here, $\rb = \rb_i -\rb_j$ is the vector connecting dipoles $i$ and $j$, whose $\alpha-$component and modulus is denoted by $\rb_\alpha$ and $r=|\rb|$, $\boldsymbol{\wp}_i^\alpha$ is  the $\alpha-$component of the vector polarization of emitter $i$, $k_0 = \omega_0 /c$ is the wave-number associated with the transition, and $\Gamma_0 = \left| \wp \right|^2 k_0 ^3 / 3 \pi \epsilon_0$ is the spontaneous emission rate of a single emitter in the ring. Throughout this work we consider the low light intensity limit, for which the observables of interest are well described by the non-Hermitian effective Hamiltonian $\Heff = \sum_{ij}  \left( J_{ij} -i\frac{\Gamma_{ij}}{2} \right) \heg_i \hge_j$. The remaining stochastic (jump operator) terms only lead to higher order corrections in the light intensity. 

\section{Sub-radiant modes in the coupled ring-impurity system}
As shown before \cite{moreno2019subradiance}, a sub-wavelength rotationally invariant ring of emitters exhibits guided modes which are extremely subradiant, but whose field vanishes at the center of the ring due to symmetry. Thus they are decoupled from any emitter at the center. Here we identify a different kind of subradiant states with large center impurity occupation dipole-dipole coupled to the symmetric bright mode of the outer ring (antenna). Such states can be interpreted as anti-resonances, where the field created by the center impurity oscillates with opposite phase and almost perfectly cancels out the effective collective field induced by pumping the outer antenna ring \cite{plankensteiner2017cavity}. In this case the large population of the central emitter in combination with the small decay rate of the coupled dark eigenmode leads to low loss energy transfer and an enhanced absorption cross-section without significant excitation of the antenna.

For concreteness we restrict ourselves to the symmetric case where all the emitters are circularly polarized in the ring plane. Nevertheless, many of the presented ideas are still valid in a more general polarization scheme. In such a symmetric configuration all emitters equally couple to the central impurity reducing the problem complexity. Specifically, in the single excitation subspace the center only couples to the single symmetric collective antenna mode. In this case the effective Hamiltonian $\Heff$ can be rewritten as 
\begin{align}
\Heff &= \Heff^I + \Heff^R + \notag \\
      &+\sqrt{N \Gamma_I/\Gamma_0} (J-i\Gamma/2) \left[ S^\dagger_{m=0} \hge_I + S_{m=0} \heg_I \right],\label{Heff_JC}
\end{align}
with  $\Heff^I = -(\delta_I + i \Gamma_I/2)\hee_I$ and $\Heff^R = \sum_{m=0}^N (J_m -i\Gamma_m/2) S^\dagger_m S_m$, being $J_m$ and $\Gamma_m$ the collective frequency shift and decay rate of the ring mode associated with the creation operator $S^\dagger_m = (1/\sqrt{N}) \sum_{j=1}^N e^ {i \pi m j/N} \heg_j$. Since only the fully symmetric mode $m=0$ couples to the impurity and thus, participates in the impurity dynamics, the relevant single ring contribution reduces to $\Heff^R = (J_R  -i\Gamma_R/2) S_{m=0}^\dagger S_{m=0}$. From now on and for notational simplicity $S^\dagger \equiv S^\dagger_{m=0}$. The ring-impurity dispersive and dissipative couplings are given by $J = \text{Re}~\mathcal{G}$ and $\Gamma = -2\text{Im}~\mathcal{G}$, with   
\begin{align}
\mathcal{G} = \frac{3\Gamma_0}{8 k_0^3 R^3} e^{i k_0 R} \left[-1 + i k_0 R + k_0^2 R^2\right].
\end{align}

For a symmetric single excitation ring mode the effective field created at the impurity position is exactly the same as that of a single dipole with enhanced dipole moment strength $\sqrt{N} \wp$ separated by a distance $R$. Thus, for a symmetric driving the system is formally equivalent to two unequal emitters with chosen parameter values. In particular the dipole replacing the ring mode is detuned from the impurity frequency transition by $J_R +\delta_I$ (see the Appendix for a more detailed study of this system). For more general, spatially varying pump fields the two systems are not totally equivalent and can exhibit different absorption cross sections.

\begin{figure}[t]
\centering
\includegraphics[width=0.49\textwidth]{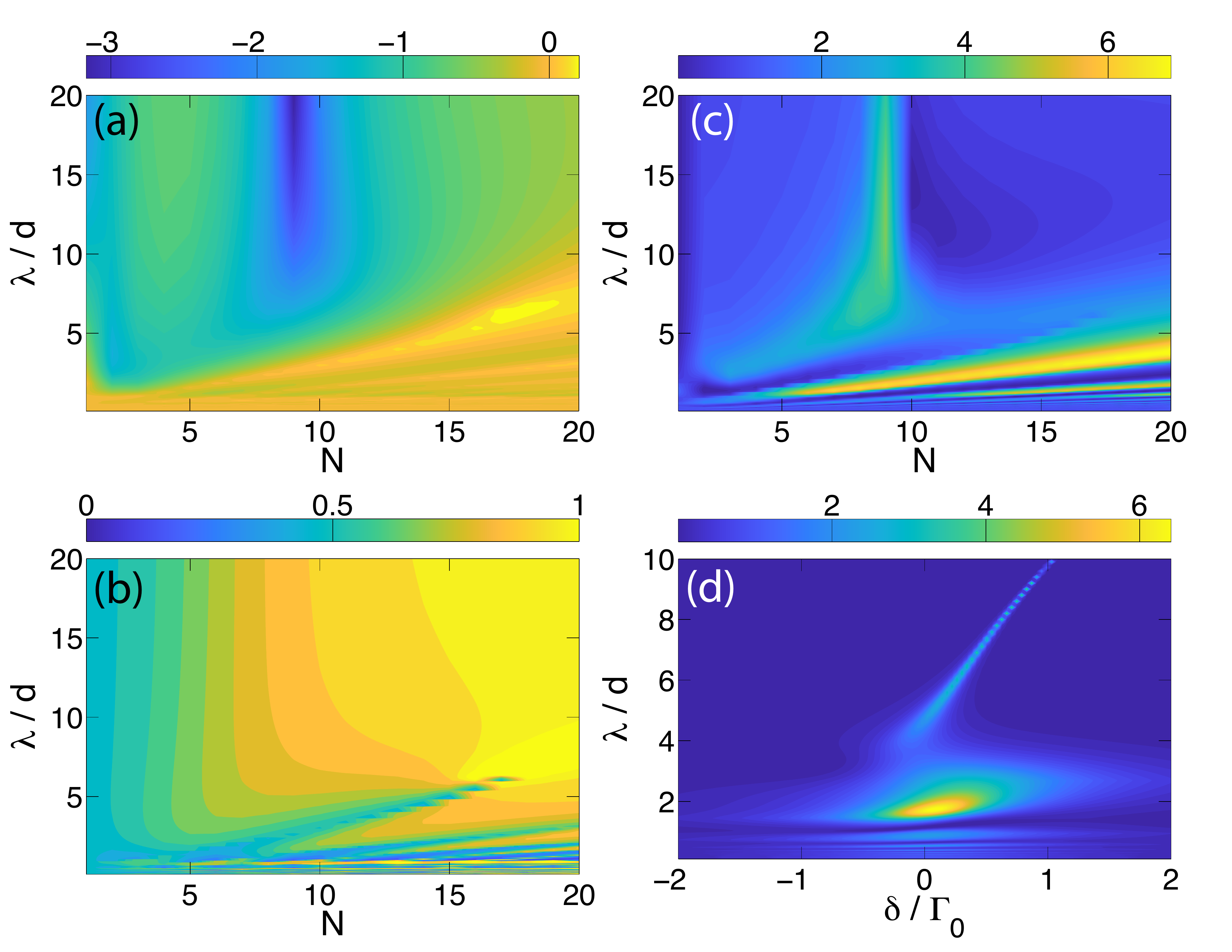}
\caption{Eigenstate properties and absorption cross section. (a) Collective decay rate (in units of $\Gamma_0$) and (b) impurity excited state population of the most subradiant eigenmode of the coupled system (ring+impurity), versus the ring atom number $N$ and light wavelength $\lambda/d$. (c) Absorption cross-section $\sabs$ (in units of maximum single atom absorption cross section $\sigma/4$) versus $N$ and $\lambda/d$, when the whole system is weakly driven by a coherent circularly polarized plane wave perpendicularly propagating (with respect to the ring plane) and on resonance with the most subradiant eigenmode. (d) For a fixed atom number in the ring $N=9$, $\sabs$ (in units of $\sigma/4$) versus external field detuning and $\lambda/d$. For a small enough ring ($\lambda/d \gg 1$) the system behaves as a frequency-selective antenna displaying a sharp resonance where the absorption cross-section is enhanced with respect to the single atom case. This frequency corresponds to the darkest eigenmode and can be tuned by changing the system paramaters.}
\label{Fig1}
\end{figure}

For the single excitation manifold we readily find the eigenmodes by diagonalizing the $2\times 2$ matrix resulting from  projecting Eq.\eqref{Heff_JC} into the subspace spanned by the basis $\left\{ \ket{R}, \ket{I} \right\}$, with $\ket{R} = S^\dagger \ket{g}$ and $\ket{I} = \heg_I \ket{g}$. The corresponding eigenvalues are 
\begin{equation} \nonumber
\lambda{\pm} = (\lambda_R + \lambda_I)/2 \pm \sqrt{(\lambda_R-\lambda_I)^2+4\Gamma_I N(J-i\Gamma/2)^2 }/2,
\end{equation} 
with $\lambda_R = J_R - i\Gamma_R/2$ and $\lambda_I = -\delta_I -i \Gamma_I/2$. In Fig.\ref{Fig1}(a) and (b) we show the decay rate and impurity occupation of the most subradiant eigenmode for the case where ring emitters and impurity are identical ($\delta_I = 0$ and $\Gamma_I = \Gamma_0$) as a function of the system geometry determined by the emitter number and ring size. As a central and surprising fact we find that in the sub-wavelength regime ($\lambda_0 / d \gtrsim 5$) an extremely dark mode with suppressed decay rate $\Gamma_{\rm min}/\Gamma_0 \lesssim 10^{-3}$ emerges exclusively when the ring contains exactly $N=9$ emitters.

The appearance of this subradiant mode can be understood with the picture of two effective dipoles in mind. Intuitively one expects that a subradiant state arises when the two dipoles have similar magnitude but opposite phase (singlet configuration), so that their radiated far fields cancel. For a generic state of the form $\ket{\Psi} = \alpha \ket{R} + \beta \ket{I}$, this implies $\beta \approx -\alpha \sqrt{N/\Gamma_I}$. In general, however, such state is not an energy eigenmode of the system. In the deep subwavelength regime ($R/\lambda \ll 1$), where $\Gamma_R \approx N\Gamma_0$ and $\Gamma \approx \Gamma_0$ it is straightforward to show that this state is only an eigenmode if we have: 
\begin{align}
J_R+\delta_I \approx J (N-\Gamma_I/\Gamma_0).
\label{Eq:deltaI_condition}
\end{align}
For identical emitters ($\delta_I = 0$ and $\Gamma_I=\Gamma_0$) this reduces to $J_R \approx (N-1)J$. Hence all emitters including the central one experience almost the same total interaction strength with all others. Indeed we find that the closest integer value $N$ satisfying this condition in the deep sub-wavelength regime is again $N=9$, as for the most subradiant mode. This can be understood as a special geometric property of the nonagon (regular polygon with $N=9$ sides), where the sum of the inverse cubic distances $1/r_{ij}^3$ to all other N-1 corners is closest to $(N-1)/R^3$, which is the scaling of the near field dipole-dipole interaction. 

\begin{figure}[t]
\centering
\includegraphics[width=0.5\textwidth]{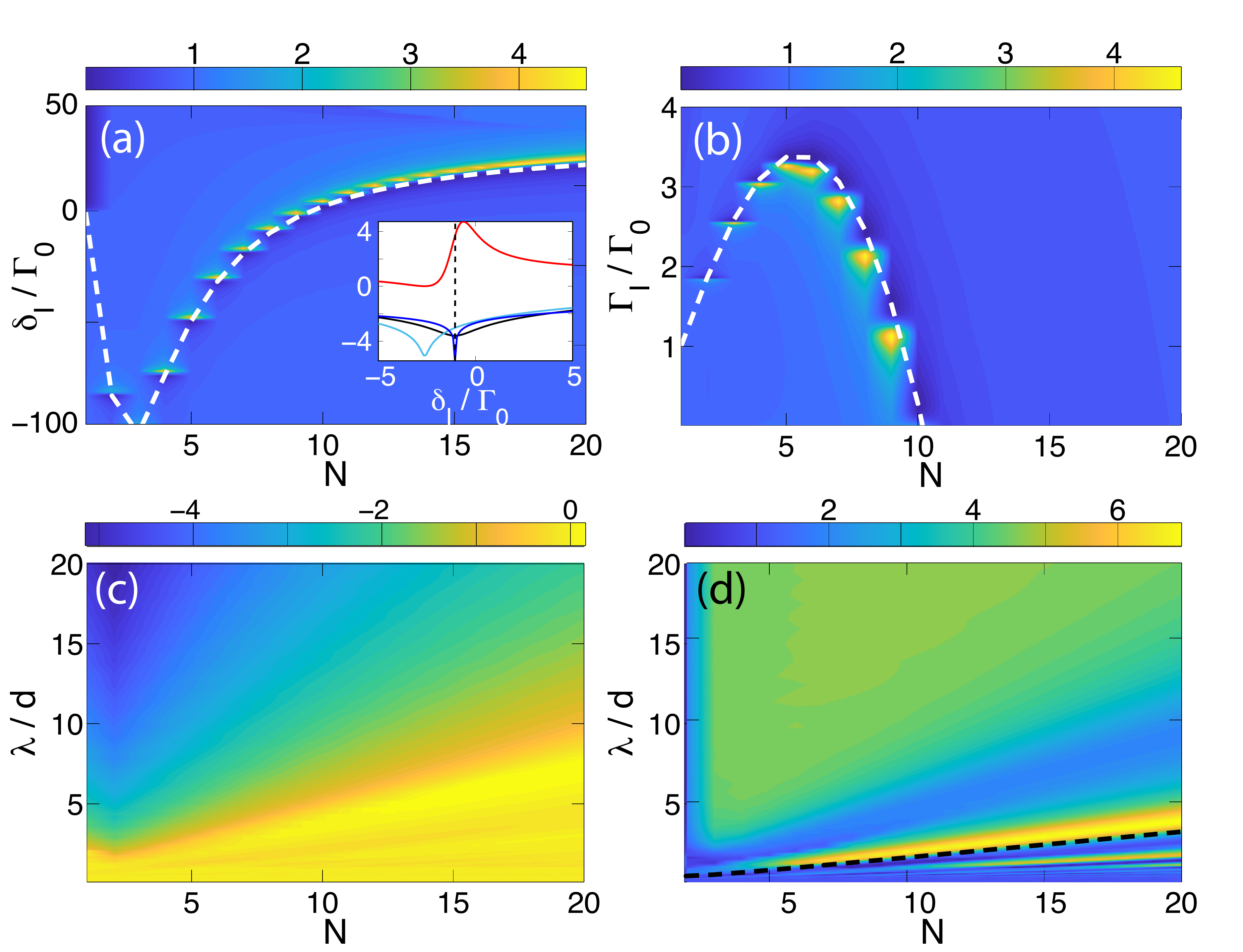}
\caption{Variation of impurity detuning $\delta_I$ and decay rate $\Gamma_I$. Absorption cross section $\sabs$ (in units of $\sigma/4$) versus $N$ and (a) $\delta_I$ at fixed $\Gamma_I = \Gamma_0$ and (b) $\Gamma_I$ at fixed $\delta_I = 0$, for $\lambda_0/d = 20$. The dashed white lines show the approximate solution given by \eqref{Eq:deltaI_condition}. The inset shows the absorption cross section for the case $N=9$ (red line), compared to the dark mode decay rate (log scale, black line). For comparison we add the effective dipole moment strength $|\wp_{\rm eff}|^2$ (cyan line) and its imaginary part (blue line, log scale). $\sabs$ is very sensitive to the detuning $\delta_I$ with a sharp maximum near the minimal imaginary part of $\wp_{\rm eff}$. (c) Decay rate (log scale) of the dark mode and (d) Absorption cross-section $\sabs$ versus $N$ and $\lambda/d$, when $\sabs$ is maximized as function of $\delta_I$ at fixed $\Gamma_I = \Gamma_0$. The dashed black line corresponds to $\lambda/R=1$. For optimal $\delta_I$ an enhanced absorption $\sabs$ is found for $N\geq 3$ at $\lambda/R \geq 1$.}
\label{Fig2}
\end{figure}

Once we have understood this general principle, also for other values of $N$ it is possible to find a similar dark mode by suitable tuning of the impurity parameters $\delta_I$ or $\Gamma_I$ to fulfill \eqref{Eq:deltaI_condition}. At finite value of $\lambda/R$ the system is very sensitive to $\delta_I$ and $\Gamma_I$ and this condition only yields an approximate solution to the optimal values. In general the  effective polarization strengths $\sqrt{N}\alpha$ and $\sqrt{\Gamma_I} \beta$ associated with the ring and impurity component of the eigenmode are complex, but the decay rate is minimized when the imaginary part is minimal and a relative phase is close to $\pi$, so that the short range interaction between the two dipoles does not contribute to the free space energy dissipation. In \fref{Fig2}(c) we show the minimum collective decay rate versus $N$ and $\lambda/d$, for the case $\Gamma_I = \Gamma_0$ and when optimizing over $\delta_I$.

\section{Absorption cross section}
Let us now study the physical consequences of this special dark resonance for light absorption. For this we add an independent decay channel at rate $\Gamma_T$ from the excited center impurity to a third auxiliary state (t) (see \fref{Fig0}). This acts as a sink extracting energy from the antenna without interaction. In light harvesting this simulates irreversible conversion of photons into chemical energy at the reaction center. In a neutral atom system one could think of a linear chain of atoms coupled to the central atom via a dipole moment component orthogonal to the antenna dipoles to irreversibly extract excitations \cite{Higgins2014Superabsorption}. For superconducting q-bits this could be a tiny antenna close to the center q-bit \cite{wang2020controllable, guimond2020unidirectional}. Mathematically we add a loss term $\mathcal{L}_T [\rho] = \Gamma_T  \left[ \hte \rho \het - (1/2) \left\{ \hee,\rho \right\} \right]$ in the master equation.

The absorption efficiency is quantified by the cross section $\sabs$ which represents the effective area for which an incident photon triggers an absorption event. Thus it gives the relative rate of absorbed photons versus incident photons, i.e. $\sabs/A = dn_{\rm abs} / dn_{\rm in}$, where $A$ is the beam area. In contrast to pure scattering or light extinction often used (e.g. \cite{cosgrove2010chlorophyll,bourne2019structure}), our definition of the absorption efficiency $\sabs$ accounts for both, the probability of scattering a photon by the system and its subsequent transfer to the auxiliary impurity state. Note that this cross section still can exceed the resonant single emitter scattering cross-section $\sigma = 6\pi/ k_0^2$. The rate of effectively absorbed photons is $dn_{\rm abs}/dt = \Gamma_T \avg{\hee_I}$, whereas the incident photon rate of an external coherent driving with Rabi frequency $\Omega$ is $dn_{\rm in}/dt = 4\Omega^2 k_0^2 A /6\pi \Gamma_0= (4\Omega^2/\Gamma_0) (A / \sigma)$, leading to $\sabs /\sigma = \Gamma_T \Gamma_0 \avg{\hee_I} / 4\Omega^2$. The system absorption efficiency will be here compared to that of a single emitter driven on resonance with spontaneous emission rate $\Gamma_0$ and including an additional decay channel at rate $\Gamma_T$. In steady state we have $\sabsc = \sigma \Gamma_0 \Gamma_T / (\Gamma_0+\Gamma_T)^2$, i.e. the product of the probabilities for first scattering a photon and subsequently absorbing it, with maximum value $\max(\sabsc ) = \sigma / 4$ for $\Gamma_T = \Gamma_0$.

For a weak coherent drive with frequency $\omega_L$ detuned by $\delta = \omega_L -\omega_0$ from the emitters resonance the steady state is given by:
\begin{align}
\ket{\Psi} = \ket{g} -i \sum_{\nu} \frac{1}{\nu - \delta} \ket{\nu} \Braket{\nu^T}{\Omega}+O(\Omega^2/\Gamma_0^2), 
\end{align}
where $\ket{\nu}$ are the eigenmodes of the system with complex eigenvalues $\nu = \omega_{\nu} - i \Gamma_{\nu}/2$ and $\ket{\Omega} \equiv \sum_i \Omega(\rb_i) \heg_i\ket{g}$. In general the excitation of dark collective modes with long lifetimes is strongly suppressed due to the inherently small overlap with propagating field modes. However, resonance enhancement of the subradiant states due to the inherent extremely small damping still yields larger absorption cross sections compared to a single emitter. For an energetically well resolved eigenmode $\ket{\nu_0}$ with a decay rate much smaller than the frequency difference to nearby eigenmodes, the steady state absorption cross section is dominated by a single term:
\begin{align}
\frac{\sabs}{\sigma} \approx \frac{\Gamma_T \Gamma_0}{\Omega^2 \Gamma_{\nu_0}^2} \left|\Braket{I}{\nu_0}\right|^2\cdot \left| \Braket{\nu_0^T}{\Omega}\right|^2.
\end{align}
It surpasses a single atom if $\Gamma_{\nu_0}/2 < \left|\Braket{I}{\nu_0}\right|\cdot \left|\Braket{\nu_0^T}{\Omega}\right|/\Omega$, i.e. the decay rate of the eigenmode has to be small but contain a large impurity occupation to compensate for the small overlap with the incoming field. 

In \fref{Fig2}(d) we show for $N=9$ (identical emitters case) versus $\lambda/d$ and detuning $\delta=\omega_L-\omega_0$ of the external coherent drive, the absorption cross section $\sabs$ in units of $\max(\sabsc ) = \sigma/4$. This shows that a narrow resonance where the absorption is greatly enhanced emerges for $\lambda/d \gtrsim 5 $, corresponding to the dark mode previously discussed. In \fref{Fig2}(b) we then plot $ \sabs $ versus $ \lambda/d$ and $N$, when the detuning of the external drive is set to the collective frequency shift of the dark mode. Clearly we observe that in the deep sub-wavelength regime ($\lambda/d \gtrsim 5$) a maximum surpassing the single emitter cross-section arises for the particular value $N=9$. Furthermore, in general, the regions with maximal absorption cross section correspond to those with minimum decay rate of the collective mode.

\begin{figure}[t]
\centering
\includegraphics[width=0.5\textwidth]{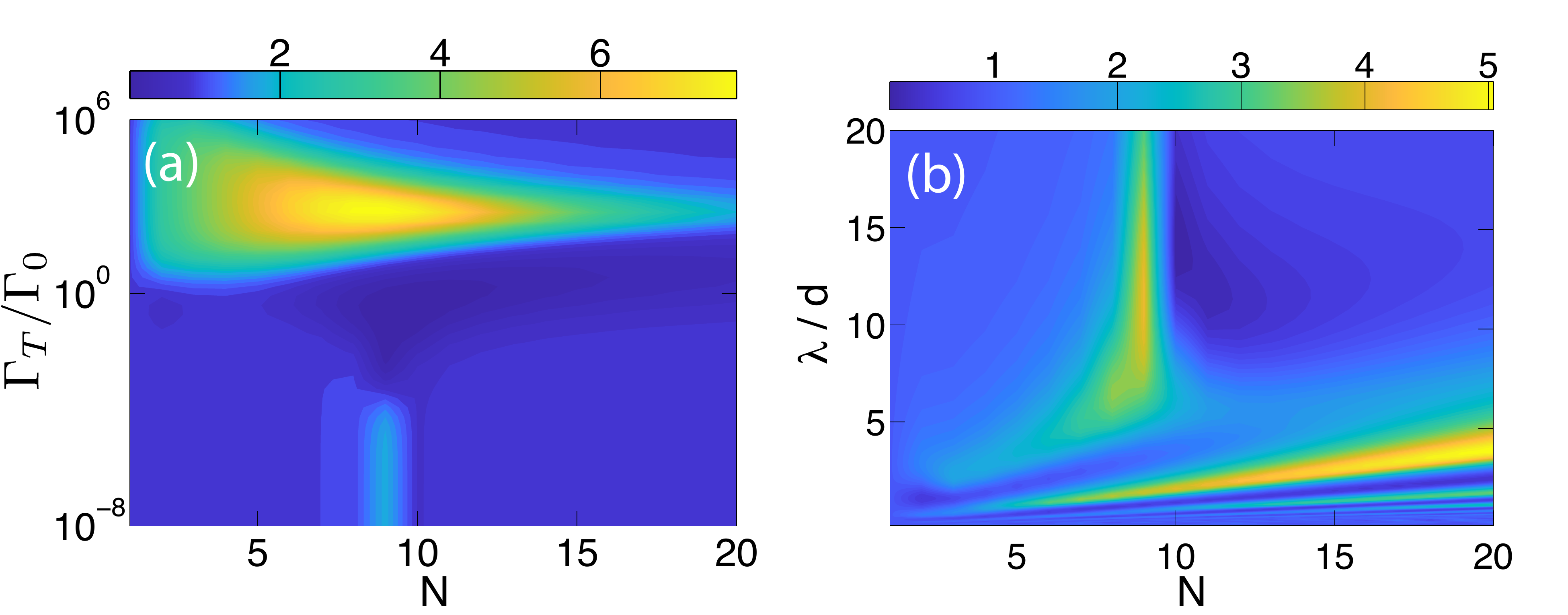}
\caption{Absorption cross-section $\sabs$ (in units of $\sabs^{\textrm{single,inc}}$ for an incoherent weak pump. (a) $\sabs$ versus $N$ and $\Gamma_T/\Gamma_0$, at fixed $\lambda/d = 40$, and (b) versus $N$ and $\lambda/d$, at fixed $\Gamma_T/\Gamma_0 = 10^{-4}$. An enhancement in absorption with respect to the single emitter case is found for $N=9$ at sufficiently small $\Gamma_T/\Gamma_0$ and large $\lambda/d$.}
\label{Fig3}
\end{figure}

As previously said a similar dark mode for different values of $N$ can be found by tuning the impurity parameters $\Gamma_I$ and $\delta_I$. Hence, for these optimal values we expect again an enhancement of the absorption cross section, as shown in \fref{Fig2}(a) and (b) where we plot $\sabs$ versus $N$ and $\delta_I$ (at fixed $\Gamma_I=\Gamma_0$), or $\Gamma_I$ (at fixed $\delta_I = 0$), respectively, for $\lambda/d = 20$. In \fref{Fig3}(d) we also plot for the optimal value of $\delta_I$, $\sabs$ as a function of $N$ and $\lambda/d$, showing that in the sub-wavelength regime an enhanced $\sabs$ with respect to the single emitter case can also be achieved for an arbitrary value of $N$ by tuning the impurity parameters.

Let us note that the results can be readily generalized to include a finite bandwidth of the incoming light. To illustrate this, we now replace the external coherent drive by a temporally incoherent (but spatially coherent) weak pump, which can be modelled by adding to the master equation the term $\mathcal{L}_{\textrm{inc}} [\rho] = \epsilon [ \hR^\dagger_{\kb} \rho \hR_{\kb} - (1/2) \{ \hR_{\kb} \hR_{\kb}^\dagger, \rho \}]$, with $\hR_{\kb} = \sum_{j=1}^{N}
e^{i \kb \rb_j} \hge_j + e^{i \kb \rb_I} \hge_I$. For a perpendicularly propagating beam of low intensity (${\kb} = 0$ and $\epsilon \ll \Gamma_0$), we show in \fref{Fig3}, $\sabs$ in units of  $\sabsinc=\sigma \Gamma_T/(\Gamma_0+\Gamma_T)$, corresponding to the absorption cross-section of a single emitter now illuminated with the same incoherent light. As it can be seen in \fref{Fig3}(a) for $\lambda/d = 40$, where $\sabs/\sabsinc $ is plotted versus $\Gamma_T$ and $N$, this ratio becomes maximal again for $N=9$ in a regime where $\Gamma_T \ll \Gamma_0$, as previously found for the coherent drive. In addition, for large enough $\lambda/d$, a broad maximum centered around $N=9$, also arises for $\Gamma_T/\Gamma_0 \sim 10^3$. In general, for a weak incoherent pump the absorption cross-section is well approximated by:
\begin{align}
\frac{\sabs}{\sigma} \approx \sum_\nu \frac{\Gamma_T}{\Gamma_{\nu}} \left|\Braket{I}{\nu} \right|^2\cdot \left|\Braket{\nu^T}{R_{\kb}}\right|^2,
\end{align}
with $\ket{R_{\kb}}\equiv \hR_{\kb}^\dagger \ket{g}$ and the index $\nu$ running over all eigenmodes. By comparing this expression to the single emitter one, we find again that the presence of a dark mode with decay rate $\Gamma_{\nu_0}$ small enough and large impurity occupation, compared to the overlap with the incident field, i.e. if $\Gamma_{\nu_0} / (\Gamma_0+\Gamma_T)< \left|\Braket{I}{\nu_0}\right|\cdot \left| \Braket{\nu_0^T}{R_{\kb}}\right|^2$, leads to an enhancement of $\sabs$. 

\section{Conclusions}

We have shown that a regular nonagon of dipoles with an additional center absorber at sub-wavelength extension, as it most commonly appears in natural light harvesting complexes, has very special geometric properties, which allow for the existence of an extremely subradiant single excitation eigenstate with a high center occupation. This particularity creates an enhanced absorption cross section with a narrow resonance at a well defined frequency. Actually, the absorption enhancement is much stronger when only the antenna dipoles are illuminated while the reaction-center is shielded. Tailoring the dipole properties of the center impurity similar enhancements can also be found for different antenna ring sizes.  
Interestingly a classical mean field description of the interacting dipoles reproduces the results only for large ring size and atom number, while the strongest enhancement at sub-wavelength distances appears only in a full quantum treatment. 

Such bio-inspired ring antenna configurations could find applications in spectroscopy as a nanoscale single photon source \cite{holzinger2020nanoscale}, detector or subtractor as higher excitations in the antenna ring are strongly suppressed by geometry. Many favorable properties survive for illumination with broadband light or including dephasing. While it is not straightforward to see how the presented enhancement effect is actually exploited in natural LHC2 light absorption, it is hard to imagine that the particular 9-fold symmetry of one of the most abundant biological structures is a pure coincidence.       

\section{Acknowledgements}
We thank David Plankensteiner, Claudiu Genes and Laurin Ostermann for helpful discussions. We are grateful to D. E. Chang for illuminating insight and ideas related to this  work. We acknowledge funding from the Austrian Science Fund (FWF) doctoral college DK-ALM W1259-N27 (R. H.), and the European Unions Horizon 2020 research and innovation program under the Marie Sklodowska-Curie grant agreement No. 801110 and the Austrian Federal Ministry of Education, Science and Research (BMBWF) (M.M.-C.). It reflects only the authors view and
the Agency is not responsible for any use that maybe
made of the information it contains. Numerical simulations were performed the Julia programming language including the QoJulia.org quantum optics package \cite{kramer2018quantumoptics}.

\bibliography{darkring}
\appendix
\section{Appendix 1: Simplified model based on an effective dipole to replace the ring}
A regular polygon of antenna emitters with a single center dipole as excitation receiver as discussed above constitutes a highly symmetric arrangement.  Hence, as it has been seen and used in previous work \cite{moreno2019subradiance}, rotation symmetry allows for an explicit calculation of eigenmodes for a simplified treatment. Nevertheless the resulting full description still is quite complex and requires a large Hilbert space. Thus the key physics is not always very easy to extract and analyze.

Here we will thus reduce the system to the minimal nontrivial size. For this we use only two effective dipoles, an antenna dipole $\mu_a  $  with a large dipole moment proportional to the square root of the number of ring atoms representing the ring  and a weak receiver dipole
$\mu_c = \mu $ at distance $R$ with an extra loss channel to represent the reaction center with corresponding energy extraction. Hence mathematically we end up with single two-level emitter for the ring and an $\Lambda$-type level system for the center.  Interestingly we will see, that still a great deal of the essential physics that we found above, can be analyzed in such a oversimplified form.

As we will deal with spatially uniform and weak excitation fields, they will almost exclusively couple to the dominant symmetric mode and excitation is limited to the single excitation manifold in the ring. Hence we replace the ring of N emitters, with a dipole moment $\mu$ and decay rate $\Gamma$ each, by a single effective antenna  with N-fold dipole moment $\mu_a = \sqrt{N_a} \mu $  placed at a small distance $R$ to the center dipole with moment
 $\mu_c=\mu$. Both dipoles are modeled as two-level systems $(\ket{0},\ket{1})$ with a circular polarized transition dipole moment and closely related transition frequency. The center atom is modeled as a three level $\Lambda$ system, where spontaneous emission from the excited level on a second independent transition towards an additional state $\ket{T}$ mimics the receivers energy absorption. The dipole-dipole interaction is then determined by two parameters characterizing the effective dipole-dipole interaction with real part  $\Omega(R)$ and imaginary part  $\Gamma(R)$.

The simplified Hamiltonian of our system thus reads:
\begin{equation} \label{Hamiltonian}
\begin{split}
H &= \hbar \omega_c \sigma_c^+ \sigma_c^- + \hbar \omega_a \sigma_a^+ \sigma_a^- \\ &+\hbar \omega_l P_l
+  \hbar \Omega(R) (\sigma_a^+ \sigma_c^- + \sigma_a^- \sigma_c^+ )
\end{split}
\end{equation}

with $\omega_l,\omega_c,\omega_a$ denoting the energies of the auxiliary loss level, center atom and antenna atom respectively and $P_l$ the Projector on the loss level.
Note that for the corresponding full geometry the respective energies and coherent shifts are given by $\omega_a = \Omega_\mathrm{sym}$ and $\Omega(R) = \sqrt{N} \Omega_{a c}$ with $\Omega_\mathrm{sym} = \sum_{i\neq c} \Omega_{i c}$ being the Eigenenergy of the symmetric state in the ring. The spontaneous decay rates of the two dipoles is then related by  $\Gamma_a = N^2 \Gamma_c $.  At very close (sub-wavelength) distance we have:\\
 $\Gamma \approx \sqrt(\Gamma_a \Gamma_c) = N \Gamma_c$ and $\Omega \ll \Gamma$.  Thus collective spontaneous decay including incoherent pumping is described by the Liouvillian

 \begin{equation} \label{Liou}
\begin{split}
        L \rho = \sum_{ij\in \{a,c\}} & \frac{\Gamma_{ij}}{2} \Big( 2 \sigma^-_i \rho \sigma^+_j - \{\sigma^+_i \sigma^-_j,\rho \} \Big) \\
        & + \frac{\Gamma_l}{2} \Big( 2 \sigma^-_l \rho \sigma^+_l - \{\sigma^+_l \sigma^-_l,\rho \} \Big) \\
        & + \frac{\nu_{ij}}{2} \Big( 2 \sigma^+_i \rho \sigma^-_j - \{\sigma^-_i \sigma^+_j,\rho \} \Big).
\end{split}
\end{equation}

 with  $ \Gamma = \begin{bmatrix} \Gamma_a & \Gamma_{ac}(R) \\ \Gamma_{ac}(R) &\Gamma_c \end{bmatrix} $ and $ \nu = \begin{bmatrix} \nu_a & \nu_{ac} \\ \nu_{ac} &\nu_c \end{bmatrix} $ . \\

The two single excitation eigenstates of $H$, denoted as
\begin{equation}
\ket{\pm} = (\ket{01} \pm \ket{10})/\sqrt{2},
\end{equation}

have energies $E_{\pm} = \hbar (\omega \pm \Omega)$. While for N=1 the energy eigenstates directly correspond to the most dark and bright superposition states with respect to decay, for $N > 1$  the states $\ket{\pm}$ are not eigenstates of the decay matrix $\Gamma$. The most bright $\ket{B}$ and dark states $\ket{D}$ are given by

\begin{align}
\ket{B} &= (c_{-} \ket{10} + \ket{01})/\mathcal{N}_-	\\  \ket{D} &= (c_{+} \ket{10} +  \ket{01})/\mathcal{N}_+ .
\end{align}
where $c_{\pm} = (\Gamma_a -\Gamma_c \mp \sqrt{\Gamma_a^2-2\Gamma_a \Gamma_c+\Gamma_c^2+4\Gamma_{ac}})/2\Gamma_{ac}$ and $\mathcal{N}_\pm$ is a normalisation constant.
Hence, for a large effective antenna ring atom number N, where $c_+ \ll 1$, the dark state $\ket{D}$ carries most excitation within the center atom. Hence it possesses a large decay rate to the target loss state $\ket{T}$ without much loss to free space. This properties play a key role in the absorption and energy loss dynamics of our coupled toy system as outlined below. \\

In order to study the light absorption in the system we can prepare it in state $\ket{G} = \ket{00}$ and simply calculate the population transfer to the final state $\ket{T}$ for various excitation an decay scenarios for a given short illumination time. Let as first start with incoherent excitation in the weak excitation regime. Here we look at three generic cases of excitation, but pumping (a) only the strong effective dipole or (b) only the center (weak) dipole and alternatively (c) collective driving of both dipoles simultaneously. Typical absorption cases are shown in Fig.\ref{FigA1}. We clearly see that increasing the dipole moment of the antenna dipole leads to much faster population accumulation in the center absorber = receiver atom. At the same time we note that for increased relative strength of the antenna the difference between collective pumping of both dipoles or only exciting the large antenna dipole diminishes.

\begin{figure}[h]
\centering
\includegraphics[width=0.45\textwidth]{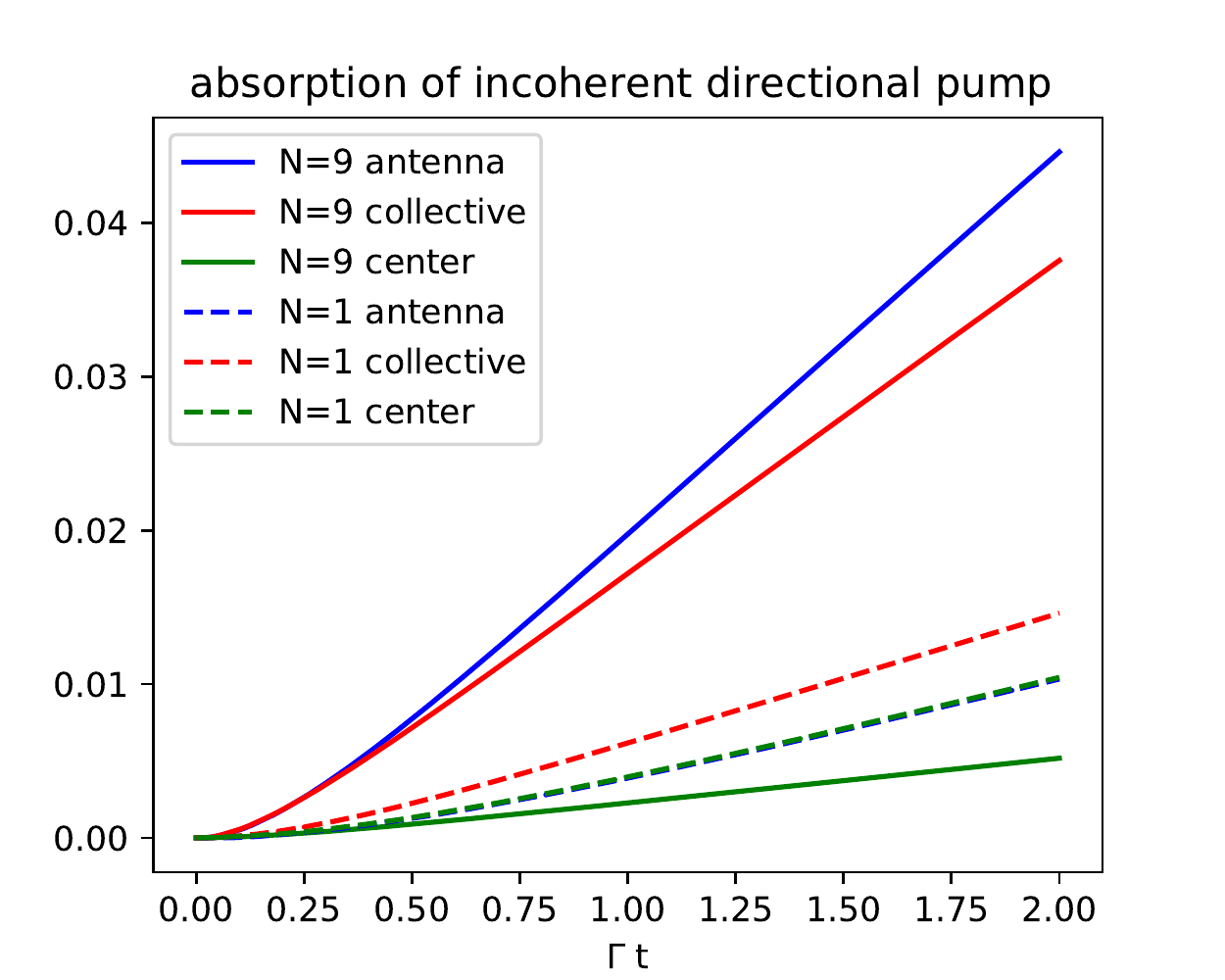}
\caption{Target state population as function of time $\Gamma t$ for incoherent collective excitation of the effective antenna atom only, the center atom only or both, where the ring is replaced by an effective single dipole of strength $ \mu_{eff} = 1* \mu$ (dashed lines) and strength $\mu_{eff} = \sqrt{9} \mu$ (solid lines). For pump and impurity loss rate we have chosen $\nu = 0.01\Gamma$ and $\Gamma_T = 3\Gamma$. }
\label{FigA1}
\end{figure}

When we relax the energy resonance condition between antenna and receiver dipole, we see pronounced differences between the three cases. In particular for selective incoherent pumping of the antenna the energy transfer to the center is resonantly enhanced at suitable receiver atom energy shift. Here the range of useful detunings gets larger with closer spacing of the receiver to the antenna. Note that as we simply assume incoherent antenna excitation the observed resonance is connected to resonant energy transfer and not to selective excitation.

\begin{figure}[h]
\centering
\includegraphics[width=0.48\textwidth]{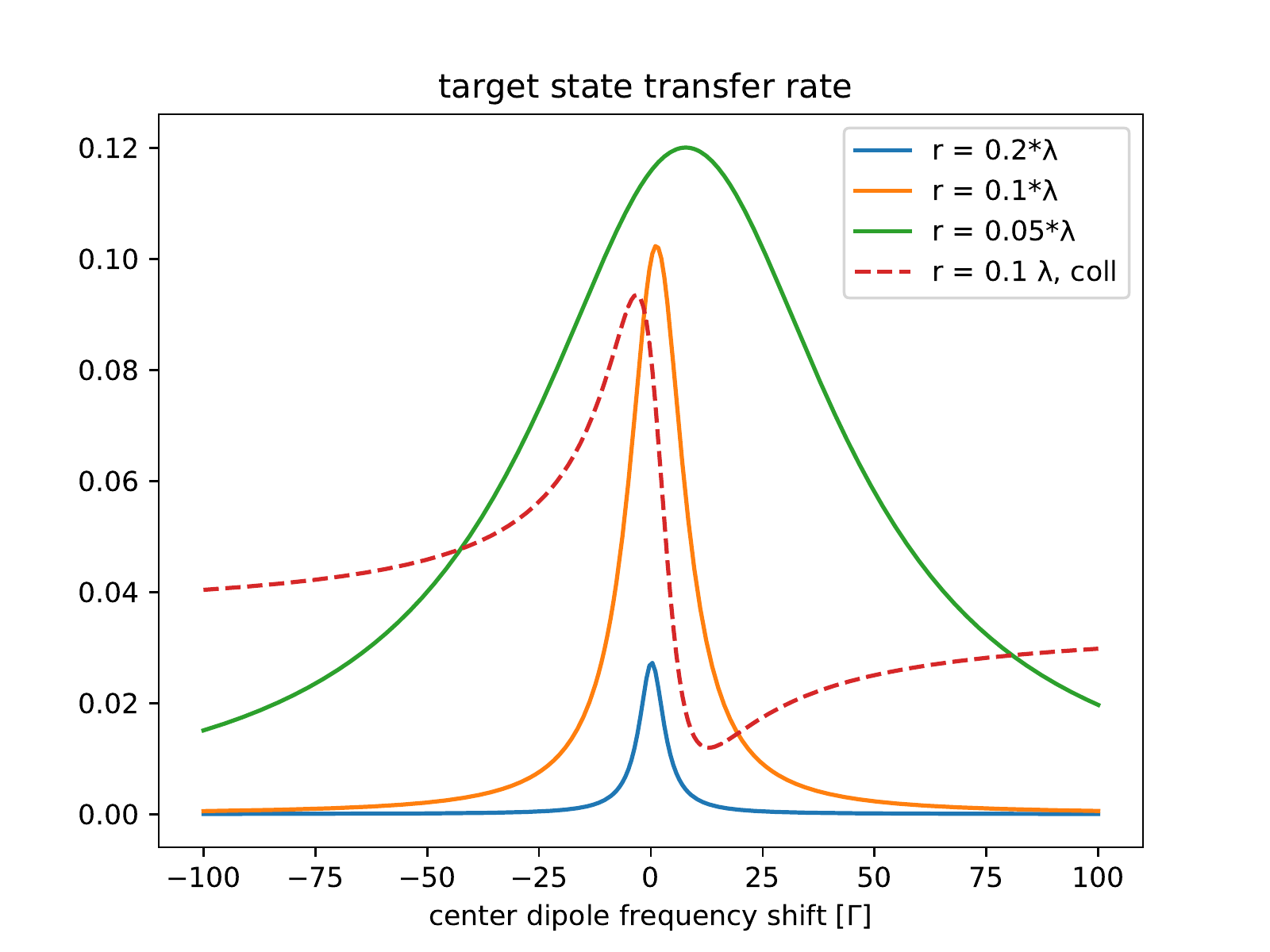}
\caption{Target state population transfer rate as function of relative center dipole resonance frequency shift for weak incoherent antenna atom excitation and effective dipole strength of $\sqrt{9} \mu$ (solid lines) for three dipole distances $r=(0.2,0.1,0.05)$. We see a strong increase of the resonant transfer for shorter antenna - receiver distances. For comparison collective pumping of both dipoles at the intermediate distance shows regions of enhancement and suppression of transfer efficiency (dashed line).  }
\label{FigA1}
\end{figure}

As expected, a more complex and interesting behavior appears for driving with a coherent field with tunable frequency. Again we study the population transferred to the center trap state after a given illumination time as function of the excitation laser frequency for different effective ring dipole strengths and the two cases of antenna or collective excitation. We see two absorption maxima at the energies of the two coupled eigenstates $\ket{\pm}$. The magnitude, splitting and width of the component increases with antenna dipole magnitude.

\begin{figure}[h]
\centering
\includegraphics[width=0.45\textwidth]{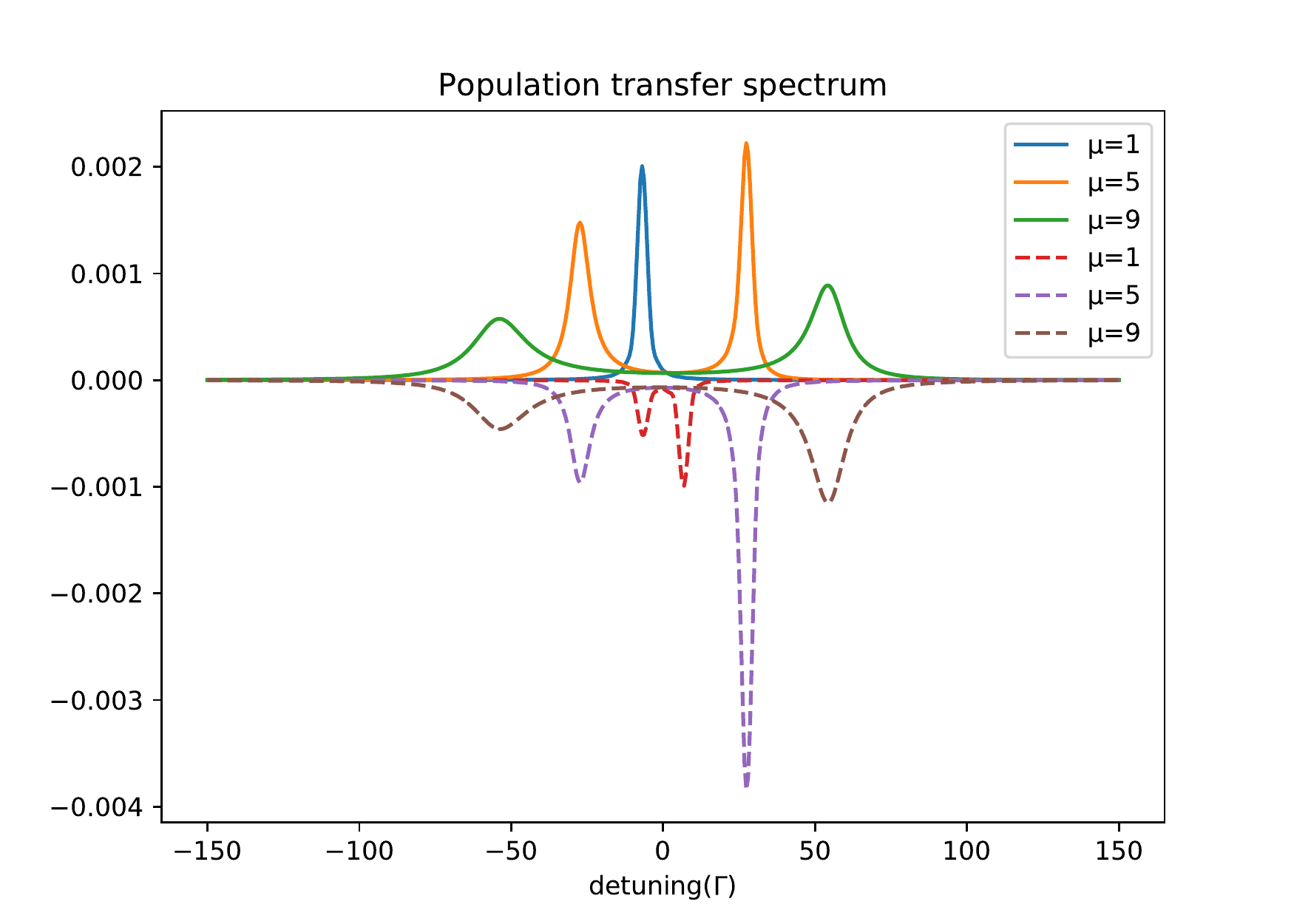}
\caption{Frequency dependence of coherent pump induced population transfer to target state for collective excitation (solid lines) and selective antenna excitation (dashed lines plotted with sign changed for better visibilty) for different effective antenna dipole moments.  }
\label{FigA2}
\end{figure}

 \begin{figure}[h]
\centering
\includegraphics[width=0.4\textwidth]{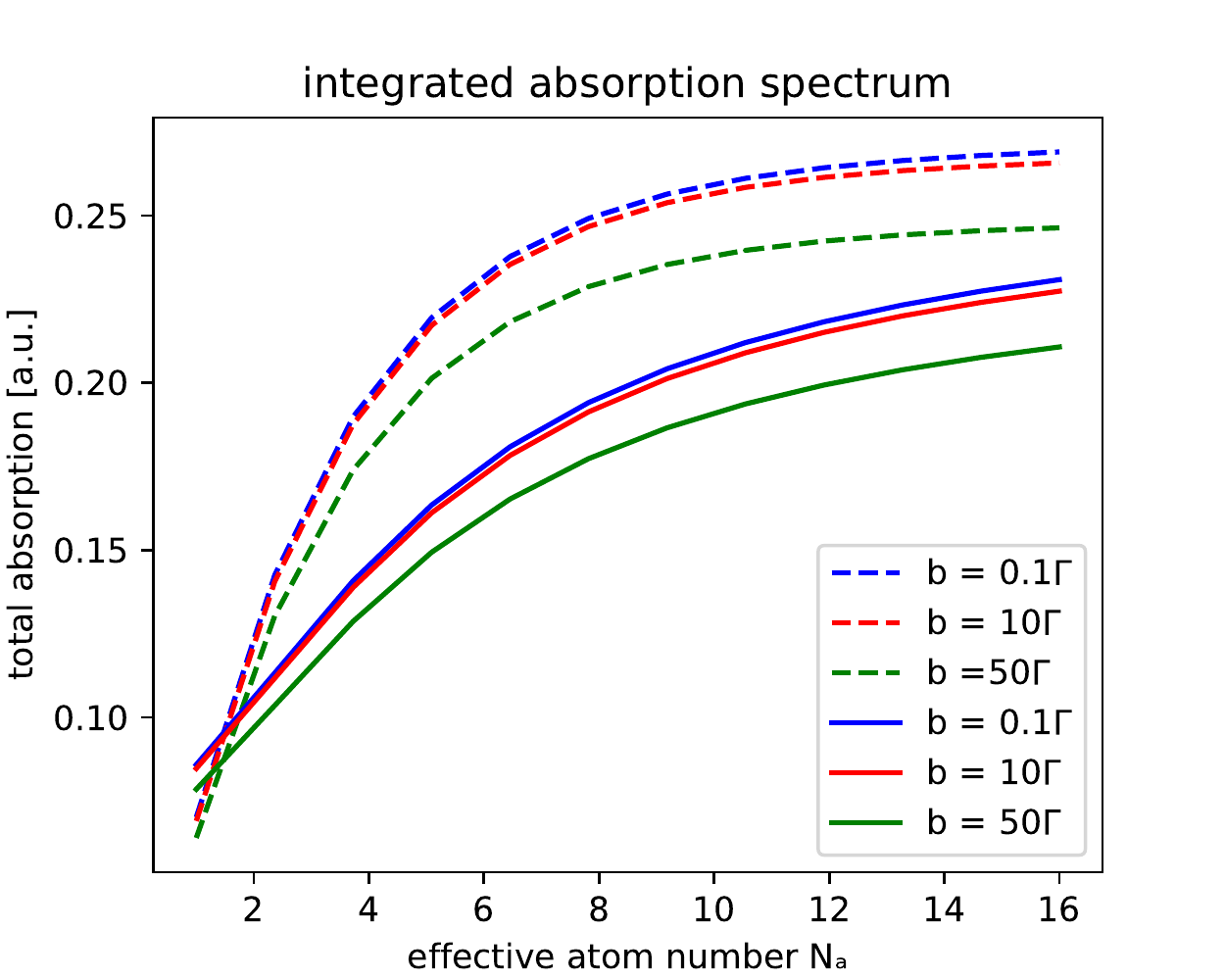}
\caption{Integrated area of absorption spectrum [a.u.] as function of antenna dipole moment $\mu$ for different laser bandwidths comparing antenna pumping (dashed lines ) and collective pumping (solid lines). We used $\Gamma_T = \Gamma$ and $\Omega = 0.1\Gamma$.  }
\label{FigA4}
\end{figure}

We see that a larger antenna dipole moment leads to a strongly broadened absorption line with a reduced maximum. Note that in course of this transfer the antenna dipole is only weakly excited and population dominantly accumulates in the dark state. Interestingly, the total area of the spectrum increases with growing antenna dipole moment up to a value of about  $N \approx 10$. Hence a nine atom ring as it appears in biological structures seem to be close to the optimum for a given amount of material.

We see that a larger antenna dipole significantly enhances the final trap state population. This enhancement is surprisingly stable against laser phase fluctuations (laser bandwidth). While for small antenna dipoles selective pumping of the antenna only clearly is favorable, the difference gets smaller for larger antenna dipoles. Again a value of $N=9$ already captures most of the enhancement and further increase of the antenna dipole only adds minor gain. Let us remark here that for collective coherent driving, secondary absorption maxima appear at larger distances of about $R\approx 0.75 \lambda$.

 \begin{figure}[h]
\centering
\includegraphics[width=0.55\textwidth]{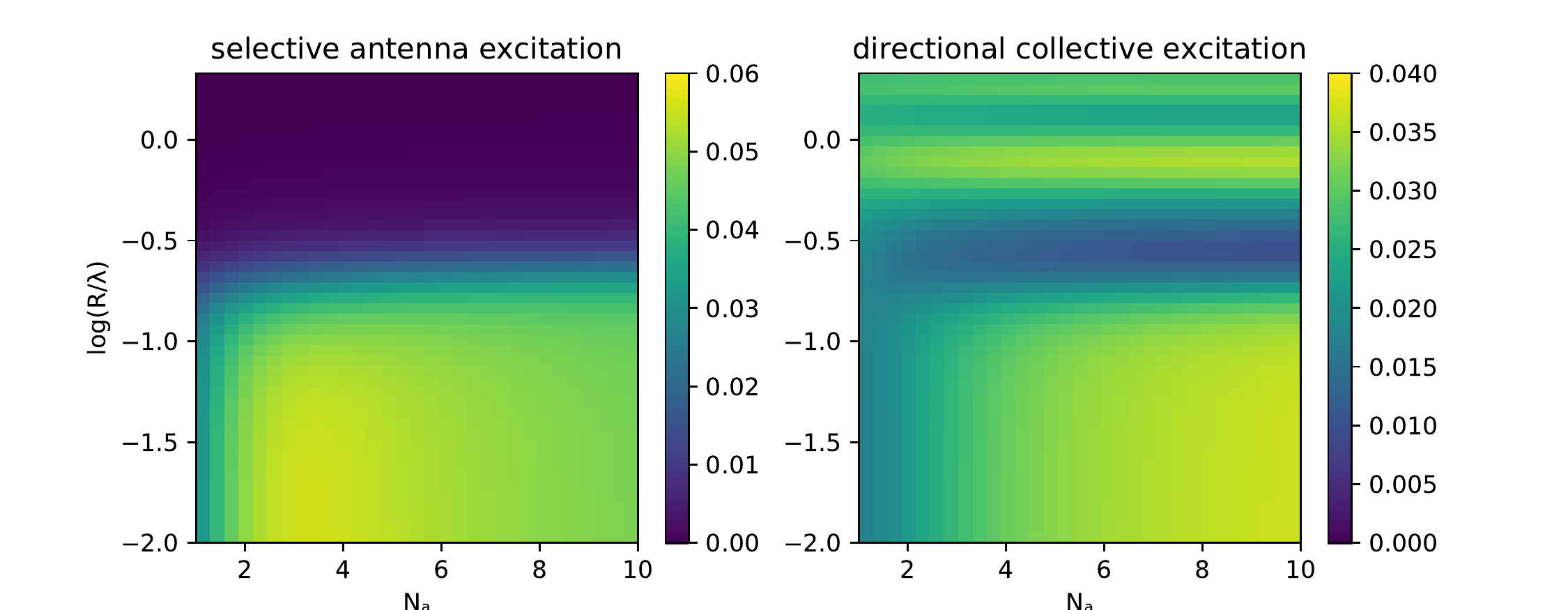}
\caption{Fraction transferred to target state after a fixed short illumination time for incoherent broadband light as function of antenna dipole moment $\mu$ and antenna to absorber distance comparing selective antenna pumping (left) and collective pumping (right) }
\label{FigA3}
\end{figure}

Finally we come back to the large bandwidth driving limit, where we can replace coherent excitation simply by transition rates. Again we compare antenna and collective excitation and vary the distance. We see that for equal excitation rates, transfer is strongly enhanced at short distances and unequal dipole moments. Similar as above, antenna only driving is generally more effective but the difference gets insignificant at about $N \approx 10$ ring dipoles.  Hence overall, splitting a light absorbing structure into a strong antenna system and a dedicated energy receiver, as also seems present in biological systems, has several generic benefits.

\section{Appendix 2: Semi-classical coupled dipole model}
\begin{figure}[t]
\includegraphics[width=0.48\textwidth]{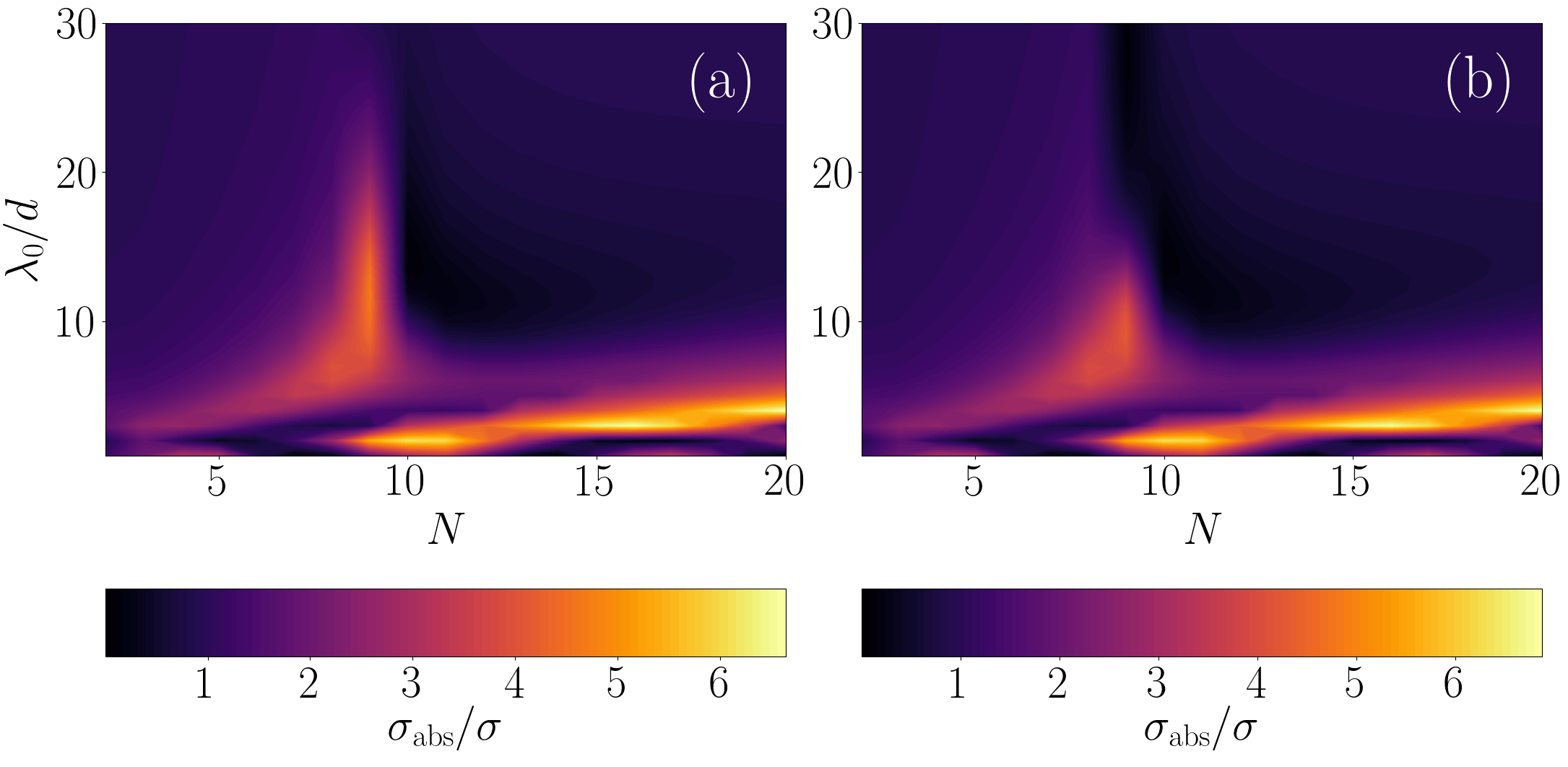} 
\caption{Effective absorption cross section as function of emitter number and emitter distance in the ring for the full quantum model (a) and the classical coupled dipole model (b). All emitters are circular polarized and the parameters are chosen as in Fig. \ref{FigA2_2}.}
\label{FigA2_1}
%\end{figure}
%\begin{figure}[h!]
\centering
\includegraphics[width=0.43\textwidth]{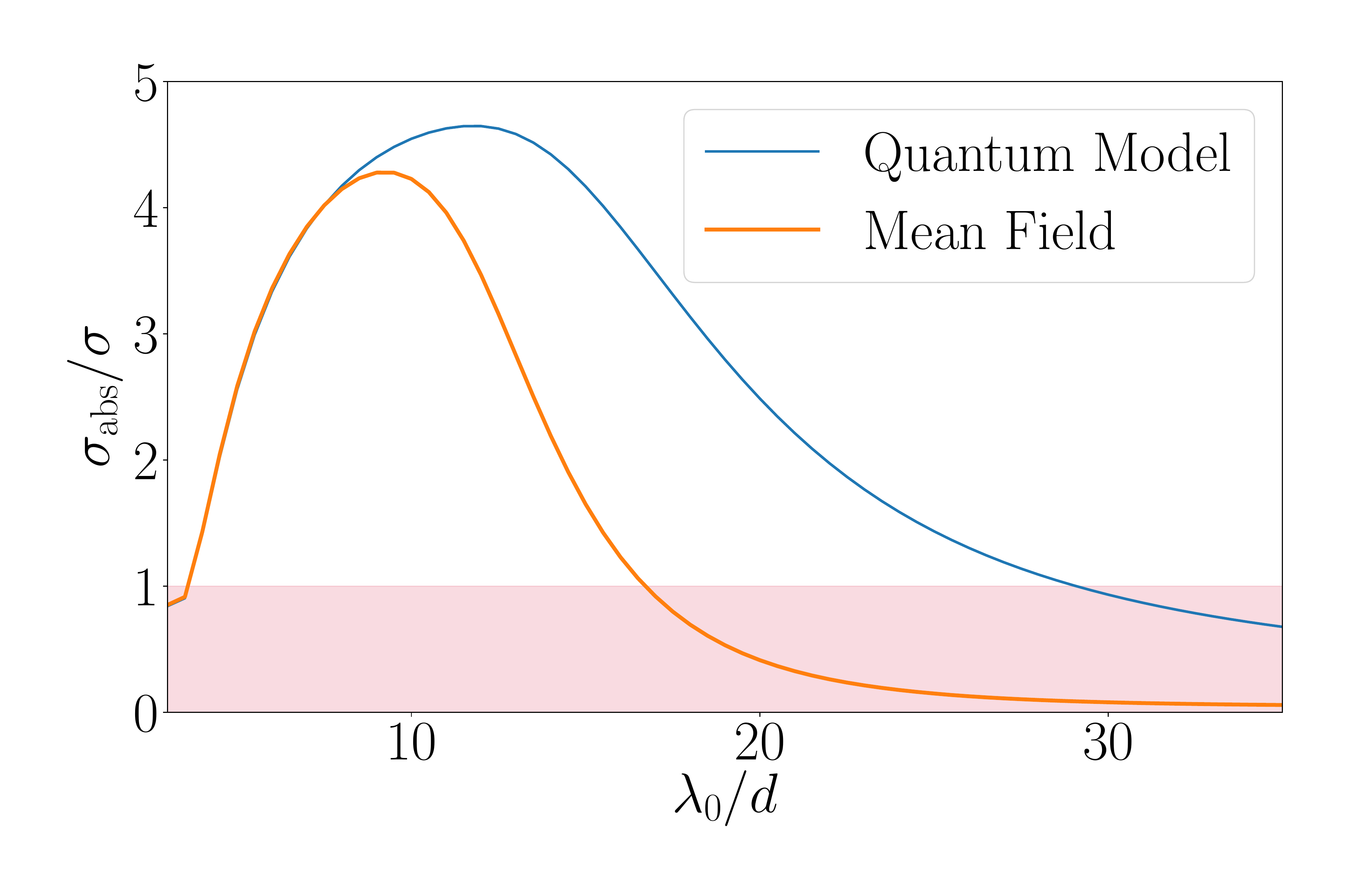}
\caption{Absorption cross section as function of the interatomic emitter distance in the $N=9$ ring for a coherent pumping rate $\Omega_r/\Gamma_0 = 5 \times 10^{-4}$ with detuning $ \Delta = -\Re \{\lambda_{Dark}\}$ and trapping rate $\Gamma_T = -2\Im \{\lambda_{Dark} \}$ for the impurity in the center.
}
\label{FigA2_2}
\end{figure}

As we have seen above the enhancement is close tied to a very dark collective eigenstate of the system with ample weight on the center dipole. As it has been argued that the most dark states are entangled \cite{plankensteiner2015selective,ruostekoski2017arrays}, one can ask whether out results hold in the case of classical dipole arrays. To this end we can simply apply a mean-field approximation to the quantum description and study the differences to the full quantum model. In this case we get simple coupled c-number Bloch equations for each dipole and the center. For symmetry reasons all ring atoms follow the same equations so that finally we end up with a rather simple set:
\begin{align}
\dot{\langle\sigma_I^{ee} \rangle} &= -2 \Omega_r \Im \{ \langle \sigma_I^{ge}\rangle\} +(i J - \Gamma/2)  \langle \sigma_I^{ge} \rangle \langle S^- \rangle^*\\
&- (i J +\Gamma/2) \langle S^- \rangle \langle \sigma_I^{ge} \rangle^*  - (\Gamma_0 + \Gamma_T) \langle \sigma_I^{ee} \rangle. \nonumber
\end{align}
\begin{align}
\dot{\langle\sigma_I^{ge} \rangle} &= -(i \Delta +\Gamma_0 +\Gamma_T/2) \langle \sigma_I^{ge}\rangle + i \Omega_r (2 \langle \sigma_I^{ee}\rangle-1) \nonumber \\
&+ 2 \langle S^-\rangle \langle \sigma_I^{ee}\rangle (i J +\Gamma/2)-\langle S^- \rangle (i J + \Gamma/2).
\end{align}
\begin{align}
\dot{\langle S^{-} \rangle} &= -(\Delta + i J_{\mathrm{sym}} + \Gamma_{\mathrm{sym}}/2)\langle S^- \rangle \\
&-N \langle \sigma_I^{ge}\rangle (i J +\Gamma/2)- i N \Omega_r. \nonumber
\end{align}
where $S^- = \sum_{j}^N \sigma^-_j$ is the collective decay operator for the ring and $J_{\mathrm{sym}} +\Gamma_{\mathrm{sym}}/2 = \sum_{j=1}^N (\Omega_{\mathrm{1j}} + \Gamma_\mathrm{1j}/2)$ with $J$ and $\Gamma$ being the coherent and dissipative coupling of the central emitter and one of the ring atoms. These can be readily solved for steady and we can then extract the effective cross section exactly as in the full quantum model above.

Interestingly we clearly see that at larger ring size and atom number both models agree quite well. However in the case of very close dipoles strong differences appear and the quantum model predicts a superior cross section. This can be even more quantitatively visualized by looking at a cut along the $N=9$ line. Note that for small diameters $\lambda /d > 8 $ in the sub-wavelength region the quantum model predicts a significantly larger cross section. \\

\end{document}